\documentclass[aps,prd,preprint,groupedaddress]{revtex4-1}
\usepackage{amsmath}
\usepackage{physics}
\usepackage{slashed}
\usepackage{multirow}
\usepackage{bm}
\usepackage{graphicx}
\usepackage{xcolor}
\usepackage{epsfig}
\begin{document}
\title{Weak Radiative Decays of Anti-triplet Bottomed Baryons in Light-Front Quark Model}
\author{Chao-Qiang Geng, Chia-Wei Liu, Zheng-Yi Wei and Jiabao Zhang}
\affiliation{School of Fundamental Physics and Mathematical Sciences, Hangzhou Institute for Advanced Study, UCAS, Hangzhou 310024, China }
\date{\today}
\begin{abstract}
We study the weak radiative decays of ${\bf B}_b \to {\bf B}_n \gamma$ with ${\bf B}_{b(n)} $
 the anti-triplet-bottom (octet) baryons in the light front quark model.
We obtain that ${\cal B}(\Lambda_b \to \Lambda \gamma)= (7.1 \pm 0.3 )\times 10 ^{-6}$, which agrees well with the current experimental value
of $(7.1 \pm 1.7 )\times 10 ^{-6}$. We predict that ${\cal B}(\Xi_b^0 \to \Xi^0 \gamma)= (1.0 \pm 0.1 )\times 10 ^{-5}$ and ${\cal B}(\Xi_b^- \to \Xi^-  \gamma) = (1.1 \pm 0.1) \times 10^{-5}$, which are consistent with the latest upper limits set by the LHCb collaboration.
In addition, we find that the $SU(3)_F$ flavor symmetry breaking effects for the modes related to the
$b\to d \gamma$ transition can be as large as $20\%$.
\end{abstract}

\maketitle

\section{Introduction}
It is known that in the weak radiative decays associated with the $b\to s\gamma$ transition, the photons are purely left-handed in the standard model~(SM) to ${\cal O}(m_s^2/m^2_b)$ precision. Clearly, a signal  of the right-handed photons in the experiment would be a smoking gun of new physics~\cite{Pati:1974yy,Mohapatra:1974gc,Mohapatra:1974hk,Senjanovic:1975rk,Senjanovic:1978ev,Mohapatra:1980yp,Lim:1981kv,Everett:2001yy,Atwood:1997zr}. 
However, as the photon polarizations  can not be measured directly at the current experimental $b$-facilities, such as LHCb, we have to analyze the cascade decays of hadrons 
to extract the  polarization information~\cite{LHCb:2020dof,LHCb:2014vnw,Geng:2021sxe,Liu:2021rvt,GarciaMartin:2019bxm}.
In addition, since the  two-body radiative decays are factorizable,
 the processes   have  a clean background for the theoretical  computation. 
  
Recently, the LHCb collaboration has reported the following decay branching ratios~\cite{LHCb:2021hfz,LHCb:2019wwi}:
\begin{equation}
{\cal B} ( \Lambda_b \to \Lambda \gamma) = (7.1 \pm 1.7)\times 10^{-6}\,,\quad \mathcal{B}(\Xi_b^-\to\Xi^-\gamma)< 1.3\times10^{-4}\,,
\end{equation}
which are  the same sizes as the charmless nonleptonic two-body decays. Remarkably, the LHCb collaboration
has also measured the lifetimes of the anti-triplet bottom baryons~(${\bf B}_b$)  with high precision~\cite{LHCb:2014chk,LHCb:2014jst}, and 
carried out a full angular analysis   for $\Lambda_b \to J/\psi \Lambda$~\cite{LHCb:2020iux}.  
These results  in the baryon decays clearly provide great opportunities to test the SM.
On the theoretical side,  the radiative bottom decays of ${\bf B}_b \to {\bf B}_n \gamma$ with ${\bf B}_{n} $ the the low-lying octet
baryons have been studied with many approaches, such as the heavy quark effective theory~\cite{Singer:1996xh}, perturbative QCD~\cite{He:2006ud}, 
$SU(3)_F$ flavor symmetry~\cite{Wang:2020wxn},
light-cone sum rule (LCSR)~\cite{Wang:2008sm,Olamaei:2021eyo}, Bethe-Salpeter equation (BSE)~\cite{Liu:2019rpm}, quark model (QM)~\cite{Chatley:1982gw,Gutsche:2013pp,Faustov:2017wbh,Faustov:2017ous}, and effective Lagrangian~\cite{PhysRevD.51.1199}. 
In this paper, we adopt the light front quark model~(LFQM), where the quark spins and the center-of-mass motions of hadrons can be treated in a consistent and fully relativistic manner, as the wave functions of the baryons are manifestly boost invariant.

The LFQM has been extensively studied in the mesonic processes~\cite{Schlumpf:1992vq,Zhang:1994ti,PhysRevD.57.5697,PhysRevD.59.114002,PhysRevD.62.074017,PhysRevD.64.114024,Geng_2003,PhysRevD.69.074025,Geng:2016pyr,Cheng:2017pcq,Chang:2019obq,Shi:2016gqt,Shen:2013oua}
 as well as the baryon semileptonic and nonleptonic ones~\cite{Zhao:2018mrg,Zhao:2018zcb,PhysRevD.98.056002,PhysRevD.99.014023,PhysRevD.100.034025,Geng:2020fng,Geng:2021nkl}. 
However, in the LFQM, the transition form factors can only be calculated in the non-timelike region. To obtain the form factors in the timelike region, where the semileptonic and nonleptonic decays occur, certain $q^2$-dependencies must be assumed, reducing the predicting power of the LFQM. In contrast, such a 
drawback does not exist in ${\bf B}_b \to {\bf B}_n \gamma$, allowing the LFQM to be tested more rigorously.

As a complement, we will also show  the results  from the $SU(3)_F$ flavor symmetry similar to those in Ref.~\cite{Wang:2020wxn}, which works well in the bottomed meson~\cite{He:1998rq,PhysRevD.64.034002,PhysRevD.69.074002,PhysRevD.93.114002,PhysRevD.50.4529,PhysRevD.52.6356,Zhou:2016jkv,PhysRevD.91.014011,PhysRevLett.75.1703} and baryon decays~\cite{Dery:2020lbc,HE201582,PhysRevD.92.036010,PhysRevD.91.115003,SINGER1996202,Wang:2017azm} as well as the charmed meson~\cite{Grossman:2012ry,PIRTSKHALAVA201281,PhysRevD.86.014014,SAVAGE1991414,PhysRevD.92.014004} and baryon decays~\cite{PhysRevD.42.1527,ALTARELLI1975277,PhysRevD.93.056008,GENG2018265,PhysRevD.97.073006,Geng:2017mxn,GENG2019214,Wang:2017azm,Wang:2019dls,Wang:2017gxe}, and compare them with our evaluations from the LFQM. 

This paper is organized as follows. In Sec.~\ref{sec:2} we present the formalisms for the decay widths and the tensor form factors. The numerical results and discussions are given in Sec.~\ref{num}. We  conclude in Sec.~\ref{con}.

\section{Formalisms}\label{sec:2}

We consider the weak radiative decays of anti-triplet bottom baryons induced by the  quark transitions of  $b \to f\gamma$ with $ f=(s,d)$.
 We will ignore the contributions from $ W $-exchange diagrams since they are suppressed by the CKM elements.
The effective Hamiltonians from the transitions are given  by~\cite{Buchalla:1995vs}
\begin{equation}
\mathcal{H}_{eff}(b\to f \gamma)=-\frac{G_F}{\sqrt{2}}\frac{e}{4\pi^2} V^*_{tf} V_{tb} C_{7\gamma}^{eff}(\mu_b) m_b[\bar{f}i\sigma^{\mu k}(1+\gamma^5)b]\epsilon_{\mu},
\end{equation}
where $\sigma^{\mu k}=\frac{i}{2}(\gamma^\mu \gamma^\nu -\gamma^\nu \gamma^\mu)k_\nu$, and $C_{7\gamma}^{eff}(\mu_b)$ corresponds to the effective Wilson coefficient at the scale of $\mu_b$ with 
$C_{7\gamma}^{eff}(5.09~\rm{GeV})= -0.303$. 
The decay amplitudes are obtained by sandwiching $ \mathcal{H}_{eff} $ with the initial and  final states
\begin{equation}\label{amp}
\mathcal{M}({\bf B}_b\to {\bf B}_n \gamma)=-\frac{G_F}{\sqrt{2}}\frac{e}{4\pi^2} V^*_{tf} V_{tb} C_{7\gamma}^{eff}(\mu_b) m_b\langle {\bf B}_n|\bar{f}i\sigma^{\mu k}(1+\gamma^5)b|{\bf B}_b\rangle \epsilon_{\mu} .
\end{equation}
The matrix elements above can be parametrized in terms of the tensor form factors, given by 
\begin{equation}\label{defff}
\begin{aligned}
\langle{\bf B}_n|\bar{f} i \sigma^{\mu k} b|{\bf B}_b\rangle&=\overline{u}_{{\bf B}_n}[f_{1}^{T V}(k^{2})(\gamma^{\mu} k^{2}-k^{\mu} \slashed{k}) / M_{{\bf B}_b}-f_{2}^{T V}(k^{2}) i \sigma^{\mu k}] u_{{\bf B}_b}, \\
\langle{\bf B}_n|\bar{f} i \sigma^{\mu k} \gamma^{5} b|{\bf B}_b\rangle&=\overline{u}_{{\bf B}_n}[f_{1}^{T A}(k^{2})(\gamma^{\mu} k^{2}-k^{\mu} \slashed{k}) / M_{{\bf B}_b}-f_{2}^{T A}(k^{2}) i \sigma^{\mu k}] \gamma^{5} u_{{\bf B}_b}.
\end{aligned}	
\end{equation}
where $u_{{\bf B}_{b(n)}}$ stands for the Dirac spinor of ${\bf B}_{b(n)}$,  $ M$ denotes the baryon mass, and $ k^\mu $ is the momentum transfer between the initial and final states. In our case of the radiative decays,  $ k^2=0 $. We will neglect the contributions associated with the form factors of
$f_{1}^{T V}(k^{2})$ and $f_{1}^{T A}(k^{2})$ 
  unless particularly noted in the rest of this paper. 
  Consequently, the decay rates are  given as 
\begin{equation}\label{decaywidth}
\Gamma({\bf B}_b \to {\bf B}_n \gamma)=\frac{\alpha_{em}}{64\pi^4}G_F^2 m_b^2 M_{{\bf B}_b}^3 
|V^*_{tf} V_{tb}|^2 (C_{7\gamma}^{eff})^2 \left(1-\frac{M_{{\bf B}_n}^2}{M_{{\bf B}_b}^2}\right)^3 (|f^{TV}_2|^2+|f^{TA}_2|^2).
\end{equation}
with $\alpha_{em}$ the fine-structure constant of the electromagnetic  interaction.

Unfortunately, to calculate the form factors of $f_{2}^{T V,TA}(k^{2})$, the baryon wave functions are required,
 which  can not be reliably obtained from the first principle due to the nonperturbative effect. 
 In this work, we  use the approach in the LFQM for the baryon wave functions, in which a baryon state 
 with momentum $ P $ and  spin $ (S,S_z) $ is expressed as~\cite{Zhang:1994ti,Schlumpf:1992vq,Geng:2020fng,Geng:2021nkl,Cheng:2004cc,Ke:2019smy,Ke:2007tg,Ke:2012wa}
 \begin{equation}
 \begin{aligned}
 |B, P, S, S_{z}\rangle=&\sum_{\lambda_{1}, \lambda_{2}, \lambda_{3}}  \int\left\{d^{3} \tilde{p}_{1}\right\}\left\{d^{3} \tilde{p}_{2}\right\}\left\{d^{3} \tilde{p}_{3}\right\} 2(2 \pi)^{3} \frac{1}{\sqrt{P^{+}}} \delta^{3}\left(\tilde{P}-\tilde{p}_{1}-\tilde{p}_{2}-\tilde{p}_{3}\right) \\
 &  \times \Psi^{S S_{z}}\left(\tilde{p}_{1}, \tilde{p}_{2}, \tilde{p}_{3}, \lambda_{1}, \lambda_{2}, \lambda_{3}\right) C^{\alpha \beta \gamma} F_{a b c}|q_{\alpha}^{a}\left(\tilde{p}_{1}, \lambda_{1}\right) q_{\beta}^{b}\left(\tilde{p}_{2}, \lambda_{2}\right) q_{\gamma}^{c}\left(\tilde{p}_{3}, \lambda_{3}\right)\rangle,
 \end{aligned}
 \end{equation}
 where  $\Psi$ represents the vertex function between the baryon and quarks, $ C^{\alpha \beta \gamma}\ (F_{a b c}) $ corresponds to the color (flavor) factor with  $ \alpha,\beta,\gamma\ (a,b,c) $ being  its indices, and $ \tilde{p}_{i}$ are the light-front 3-momenta of the $i$-th constituent quark, defined by
 \begin{equation}
 p_i=(p_i^-,p_i^+,p_i^1,p_i^2)=(p_i^-,\tilde{p}_i)=(p_i^-,p_i^+,p_{i\perp}),
 \end{equation}
 with $ p_i^{\pm}=p_i^0\pm p_i^3 $ and $ p_i^-p_i^+-p^2_{i\perp}=m_i^2 $. Here,
the integration measure and  Delta function are given as 
 \begin{equation}
 \dd^{3} \tilde{p}_{i} \equiv \frac{d p_{i}^{+} d^{2} p_{i \perp}}{2(2 \pi)^{3}}, \quad \delta^{3}(\tilde{p})=\delta\left(p^{+}\right) \delta^{2}\left(p_{\perp}\right).
 \end{equation}
 respectively, along with the normalization
 \begin{eqnarray}
 \braket{q_{\alpha^\prime}^{a^\prime}(\tilde{p}_i^\prime, \lambda^\prime)}{q_{\alpha}^{a}(\tilde{p}_i, \lambda)}=2(2 \pi)^{3} \delta^{3}\left(\tilde{p}^{\prime}_i-\tilde{p}_i\right) \delta_{\lambda^{\prime} \lambda} \delta_{\alpha^{\prime} \alpha} \delta^{a^{\prime} a}\,,\nonumber\\
  \left\langle\mathbf{B}, P^{\prime}, S^{\prime}, S_{z}^{\prime} \mid \mathbf{B}, P, S, S_{z}\right\rangle=2(2 \pi)^{3} P^{+} \delta^{3}(\tilde{P}^{\prime}-\tilde{P}) \delta_{S_{z}^{\prime} S_{z}}.
 \end{eqnarray}
  The vertex function  can be  further decomposed as~\cite{Zhang:1994ti,Schlumpf:1992vq,Geng:1997ws}
 \begin{equation}
 \Psi^{S S_{z}}\left(\tilde{p}_{1}, \tilde{p}_{2}, \tilde{p}_{3}, \lambda_{1}, \lambda_{2}, \lambda_{3}\right)=\Phi\left(\tilde{p}_{1}, \tilde{p}_{2}, \tilde{p}_{3}\right) \Xi^{S S_{z}}\left(\lambda_{1}, \lambda_{2}, \lambda_{3}\right),
 \end{equation}
 in which $ \Phi $ is the momentum distribution function, $ \Xi^{S S_{z}} $ stands for  the helicity wave function given  as
 \begin{equation}
 \Xi^{S S_{z}}\left(\lambda_{1}, \lambda_{2}, \lambda_{3}\right)=\sum_{s_{1}, s_{2}, s_{3}}\prod_{i=1}^3\langle\lambda_{i}|R_{i}^{\dagger}|s_{i}\rangle \braket{\frac{1}{2} s_{1}, \frac{1}{2} s_{2}, \frac{1}{2} s_{3}}{S S_{z}},
 \end{equation}
 where
 $ R_i $ is the Melosh matrix, which brings the $ i $-th quark from its spin state to a helicity state,  and $ \braket{\frac{1}{2} s_{1}, \frac{1}{2} s_{2}, \frac{1}{2} s_{3}}{S S_{z}} $ is the Clebsch-Gordan coefficient, embodied in the spin wave function.

 The explicit forms of $ \Phi $ and $ R_i $ depend on the parametrization scheme of the internal motions of the constituent quarks. 
 In our calculation, we choose the diquark scheme with the first two quarks being coupled to each other, 
 while the other coupling schemes can be easily done by the permutations~\cite{Geng:2020fng}.  The kinematic variables are given as 
  \begin{equation}
 \begin{aligned}
 \tilde{P}&=\tilde{p}_{1}+\tilde{p}_{2}+\tilde{p}_{3},\qquad \xi_3=\frac{p_1^+}{p_1^++p_2^+},\qquad \eta_3=1-\frac{p_3^+}{P^+},\\
 q_{3\perp}&=(1-\xi_3)p_{1\perp}-\xi_3 p_{2\perp},\quad Q_{3\perp}=(1-\eta_3)(p_{1\perp}+p_{2\perp})-\eta_3 p_{3\perp}.\\
 \end{aligned}
 \end{equation} 
 Note that $(\xi_3 , q_{3\perp})$ describe the internal motion within the diquark system, 
 while $(\eta_3 , Q_{3\perp})$  the relative motion between the diquark and the third quark~\cite{Bakker:1979eg}. 
 The invariant masses are then given as
 \begin{equation}
 \begin{aligned}
 M_3^2&=\frac{q_{3\perp}^2}{\xi_3(1-\xi_3)}+\frac{m_1^2}{\xi_3}+\frac{m_2^2}{1-\xi_3},\\
 M^2&=\frac{Q_{3\perp}^2}{\eta_3(1-\eta_3)}+\frac{M_3^2}{\eta_3}+\frac{m_3^2}{1-\eta_3}.\\
 \end{aligned}
 \end{equation}
 
 In this work, we adopt the Gaussian-type momentum wave function for the ground state baryon~\cite{Schlumpf:1992vq,Geng:2020fng,Geng:2021nkl}. In this particular set of kinematic variables, we have that
 \begin{equation}\label{diquark}
 \phi_3\equiv \Phi(\xi_3, q_{3\perp}, \eta_3, Q_{3\perp})=\mathcal{N}\sqrt{\frac{\partial q_{3z}}{\partial\xi_3}\frac{\partial Q_{3z}}{\partial\eta_3}}\exp(-\frac{\vec{Q}_3^2}{2\beta_Q^2}-\frac{\vec{q}_3\!^{2}}{2\beta_{qq'}^2}),
 \end{equation}
 with $ \mathcal{N}=\left(\beta_{qq'} \beta_{Q} \pi\right)^{-3 / 2} $ and
 \begin{equation}
 \begin{aligned}
 q_{3z}&=\frac{\xi_3 M_3}{2}-\frac{m_1^2+q_{3\perp}^2}{2\xi_3 M_3},\quad\,\vec{q}_3\!^{2}=q_{3\perp}^2+q_{3z}^2,\\
 Q_{3z}&=\frac{\eta_3 M}{2}-\frac{m_3^2+Q_{3\perp}^2}{2\eta_3 M},\quad\vec{Q}_3^2=Q_{3\perp}^2+Q_{3z}^2,\\
 \end{aligned}
 \end{equation}
where $ \beta_{qq'}$ and $\beta_{Q} $ are the confinement energy scales within the diquak system and between the diquark and  third quark, respectively. 
Note that we take the shape parameters as the internal kinematic freedoms to describe the diquark systems
 instead of  the diquark masses.

 If the integration variables  $ (\vec{q},\vec{Q}) $ are used instead of $(\xi_3,q_{3\perp},\eta_3,Q_{3\perp}) $, we obtain that 
 \begin{equation}
 \int \dd \xi_3 \dd \eta_3 \dd^{2} q_{3\perp} \dd^{2} Q_{3\perp}\left|\phi_{3}\right|^{2}= \int \dd^3 \vec{q}_3 \dd^3 \vec{Q}_3 {\cal N}^2 \exp(-\frac{\vec{Q}_3^2}{2\beta_Q^2}-\frac{\vec{q}_3\!^{2}}{2\beta_{qq'}^2}) =1,
 \end{equation}
 where the wave functions are clearly  Gaussian.
 On the other hand, the angular dependencies are embodied in $R_i$, given as 
 \begin{equation}
 \begin{array}{l}
 R_{1}=R_{M}\left(\eta_3, Q_{3\perp}, M_{3}, M\right) R_{M}\left(\xi_3, q_{3\perp}, m_{1}, M_{3}\right), \\
 R_{2}=R_{M}\left(\eta_3, Q_{3\perp}, M_{3}, M\right) R_{M}\left(1-\xi_3,-q_{3\perp}, m_{2}, M_{3}\right), \\
 R_{3}=R_{M}\left(1-\eta_3,-Q_{3\perp}, m_{3}, M\right),
 \end{array}
 \end{equation}
 with
 \begin{equation}
 R_{M}\left(\xi, q_{\perp}, m, M\right)=\frac{m+\xi M-i \vec{\sigma} \cdot(\vec{n} \times \vec{q})}{\sqrt{(m+\xi M)^{2}+q_{\perp}^{2}}},
 \end{equation}
 where $ \vec{\sigma}=(\sigma^1,\sigma^2,\sigma^3) $, representing the Pauli matrices, and $ \vec{n}=(0,0,1) $. We emphasize that the explicit forms of $R_i$ depend on the parametrization schemes.

As the baryon wave functions are given, we are now ready to calculate the form factors.
To illustrate the calculation,
we take $\Lambda_b\to \Lambda\gamma$ as an example, while the others can be obtained with slight modifications.
 The relevant ones in Eq.~\eqref{decaywidth} can be extracted through the following equalities,

\begin{equation}\label{extract}
	\begin{aligned}
		&f_{2}^{TV}=\frac{1}{4P^{+}} \langle{\Lambda,P^\prime, \uparrow}|\bar{f}i\sigma^{R+}b|{\Lambda_b,P,\downarrow}\rangle,\\
		&f_{2}^{TA}=-\frac{1}{4P^{+}}\langle{\Lambda,P^\prime, \uparrow}|\bar{f}i\sigma^{R+}\gamma_{5}b |{\Lambda_b,P,\downarrow}\rangle,
	\end{aligned}
\end{equation}
where $ \gamma^R=\gamma^{1}+i\gamma^2,\gamma^+=\gamma^0+\gamma^3 $, and the Dirac spinors in the light-front formalism can be found in Appendix~\ref{appB}. We choose $ k^+=0 $  to perform the calculation. In the LFQMs, this particular  frame is often used to avoid the zero-mode graphs~\cite{Choi:1998nf,Choi:2005fj,Choi:2011xm,Choi:2012zzb,Choi:2013ira}. It has been shown that their contributions to the vector form factors vanish at the limit of $k^+\to 0$~\cite{Schlumpf:1992vq,Zhang:1994ti}. 
In this work, we would take it as a working assumption for the tensor ones and test it with the experimental data. 

The full wave functions of $\Lambda_b$ and $\Lambda$ are given as 
\begin{equation}
	\begin{aligned}
		\ket{\Lambda_b}&=\frac{1}{\sqrt{6}}\left[\phi_{3} \chi^{\rho 3}(\ket{udb}-\ket{dub})+\phi_{2} \chi^{\rho 2}(\ket{ubd}-\ket{dbu})+\phi_{1} \chi^{\rho 1}(\ket{bud}-\ket{bdu})\right],\\
		\ket{\Lambda}&=\frac{1}{\sqrt{6}}\left[\phi_{3} \chi^{\rho 3}(\ket{uds}-\ket{dus})+\phi_{2} \chi^{\rho 2}(\ket{usd}-\ket{dsu})+\phi_{1} \chi^{\rho 1}(\ket{sud}-\ket{sdu})\right],\\
	\end{aligned}
\end{equation}
while the others can be found in Appendix~\ref{appC}. 
In our study, since  diquark clusters are viewed as  effective particles, they are chosen in a way to acquire definite angular momenta. For $\Lambda$ and $\Sigma^0$, the $(u,d)$ pairs would  form the states with $J=0$ and $J=1$, respectively, whereas the $(u,s)$ and $(d,s)$ pairs would be the mixtures of $J=0$ and $J=1$. Thus, we choose the $(u,d)$ pairs to form diquark clusters instead of the others. This way of constructing the baryon wave functions would break the $SU(3)$ flavor symmetry by hand, as it does not allow a diquark cluster with a light quark and a strange quark inside $ \Lambda $ or $ \Sigma $. Such breaking effects are embedded in the expressions of the baryon octet, as the $\Lambda$ and $\Sigma$ baryons are taken to preserve the isospin symmetry instead of the $U-$spin or $V-$spin symmetry. 

There are 6 terms, which contribute to the transition, read as
\begin{equation}\label{perm}
	\begin{aligned}
		&\ket{udb}\to \ket{uds},~\ket{dbu}\to \ket{dsu},~\ket{bud}\to \ket{sud},\\
		&\ket{dub}\to \ket{dus},~\ket{ubd}\to \ket{usd},~\ket{bdu}\to \ket{sdu}.\\
	\end{aligned}
\end{equation}

The first one in Eq.~\eqref{perm} contributes to the form factors as
\begin{equation}
	\begin{aligned}
		(f_{2}^{TV})_{udb\to uds}=&\frac{1}{4P^{+}}\int\dd{\xi_3}\dd{\eta_3}\dd^2{q_{3\perp}}\dd^2{Q_{3\perp}}\phi_3^\prime\phi_3 F_{uds}F_{udb}\\
		&\times\sum_{\lambda_{i}^\prime, \lambda_{i}}\Xi^{\frac{1}{2},+\frac{1}{2}}(\lambda_{i}^\prime)^\dagger\delta_{\lambda_1^{\prime} \lambda_1}\delta_{\lambda_2^{\prime} \lambda_2}(O_{3}^{TV})_{\lambda_{3}^{\prime}\lambda_{3}}\Xi^{\frac{1}{2} ,-\frac{1}{2}}(\lambda_{i}),
	\end{aligned}
\end{equation}
with 
\begin{equation}\label{O3T}
\begin{aligned}
&(O_{3}^{TV})_{\lambda_{3}^{\prime}\lambda_{3}}=\frac{1}{1-\eta} \bar{u}\left(\vec{p}_{3}\,^{\prime} \lambda_{3}^{\prime}\right) i\sigma^{R+} u\left(\vec{p}_{3} \lambda_{3}\right)=-4P^{+}(\sigma^R)_{\lambda_{3}^{\prime}\lambda_{3}},\\
&(O_{3}^{TA})_{\lambda_{3}^{\prime}\lambda_{3}}=\frac{1}{1-\eta} \bar{u}\left(\vec{p}_{3}\,^{\prime} \lambda_{3}^{\prime}\right) i\sigma^{R+}\gamma_{5} u\left(\vec{p}_{3} \lambda_{3}\right)=4P^{+}(\sigma^R)_{\lambda_{3}^{\prime}\lambda_{3}},\\
\end{aligned}
\end{equation}
where $ F_{uds}=F_{udb}=1/\sqrt{6} $, and $O_3^{TV,TA}$ describe the $b\to s $ transition  at the quark level, note that we have noramlized $ \sigma^R$ as $(\sigma^1+i\sigma^2)/2 $.

On the other hand, the helicity wave functions for the initial and final baryons are
\begin{equation}
	\Xi^{\frac{1}{2},+\frac{1}{2}}(\lambda_{i}^\prime)^\dagger=\sum_{\chi^{\rho3}_\uparrow}\prod_{i=1}^3\bra{s_{i}^\prime}R_{i}^\prime\ket{\lambda_{i}^\prime},\quad\Xi^{\frac{1}{2} ,-\frac{1}{2}}(\lambda_{i})=\sum_{\chi^{\rho3}_\downarrow}\prod_{i=1}^3\bra{\lambda_{i}}R_{i}^\dagger\ket{s_{i}}.
\end{equation}
Combing them together, we get
\begin{equation}\label{FTV}
		\begin{aligned}
			(f_{2}^{TV})_{udb\to uds}&=-\frac{1}{6}\int\dd{\xi_3}\dd{\eta_3}\dd^2{q_{3\perp}}\dd^2{Q_{3\perp}}\phi_3^\prime\phi_3\\
			\times&\sum_{\chi^{\rho3}_\uparrow,\chi^{\rho3}_\downarrow}\prod_{i=1,2}\bra{s_{i}^\prime}R_{i}^\prime\cdot R_{i}^\dagger\ket{s_{i}}\bra{s_{3}^\prime}R_{3}^\prime\cdot\sigma^R\cdot R_{3}^\dagger\ket{s_{3}}.
		\end{aligned}
\end{equation}
 Similarly, $f_2^{TA}$ can be obtained, given as
\begin{equation}\label{FTA}
	\begin{aligned}
		(f_{2}^{TA})_{udb\to uds}&=-\frac{1}{6}\int\dd{\xi_3}\dd{\eta_3}\dd^2{q_{3\perp}}\dd^2{Q_{3\perp}}\phi_3^\prime\phi_3\\
		\times&\sum_{\chi^{\rho3}_\uparrow,\chi^{\rho3}_\downarrow}\prod_{i=1,2}\bra{s_{i}^\prime}R_{i}^\prime\cdot R_{i}^\dagger\ket{s_{i}}\bra{s_{3}^\prime}R_{3}^\prime\cdot\sigma^R\cdot R_{3}^\dagger\ket{s_{3}}.
	\end{aligned}
\end{equation}

By the permutation symmetry of the Fermi statistics, it is straightforward to see that the six transitions give exactly  the same contributions  to the form factors. 
As a result, the transition form factors between the baryons can be obtained by multiplying Eqs.~\eqref{FTV} and \eqref{FTA} by 6, given by
\begin{equation}\label{f2g2}
	\begin{aligned}
		(f_{2}^{TV})_{\Lambda_b \to \Lambda}=(f_{2}^{TA})_{\Lambda_b \to \Lambda}&=-\int\dd{\xi_3}\dd{\eta_3}\dd^2{q_{3\perp}}\dd^2{Q_{3\perp}}\phi_3^\prime\phi_3\\&\times\sum_{\chi^{\rho3}_\uparrow,\chi^{\rho3}_\downarrow}\prod_{i=1,2}\bra{s_{i}^\prime}R_{i}^\prime\cdot R_{i}^\dagger\ket{s_{i}}\bra{s_{3}^\prime}R_{3}^\prime\cdot\sigma^R\cdot R_{3}^\dagger\ket{s_{3}},\\
	\end{aligned}
\end{equation}
It is worth to mention that the equality 
\begin{equation}\label{feq}
 f_{2}^{TV}=f_{2}^{TA}
\end{equation}
 only holds at $k^2=0$. This result is also consistent with that in 
Refs.~\cite{Gutsche:2013pp,Wang:2008sm}. 

The equivalence can be understood intuitively in terms of the valence quark framework, in which the spin direction of $\Lambda$  is attributed to its strange quark solely. As a result, it is necessary for $\Lambda$ to have the same helicity with the strange quark, which is left-handed. A direct consequence is that 
\begin{equation}\label{equality}
H_+ = 0\,,
\end{equation}
where $H_{+(-)}$ corresponds to  the helicity amplitude with the subscript denotes the helicity of  $\Lambda$. 

Without carrying out the numerical detail, Eq.~\eqref{equality} is sufficient for one to analyze the decay angular distributions. For $\Lambda_b \to \Lambda(\to p\pi^-) \gamma$, the angular dependency is given as~\cite{GarciaMartin:2019bxm,Gutsche:2013pp}
\begin{equation}
{\cal D}(\Omega) \equiv 
\frac{1}{8\pi}\frac{1}{ \Gamma}\frac{\partial^3 \Gamma}{\partial\cos \theta\partial \cos \theta_1\partial\phi}
= 1-P_b\alpha_\Lambda\cos\theta\cos \theta_1  - P_b P_L \cos\theta + \alpha_\Lambda P_L \cos \theta_1\,, 
\end{equation}
where $P_b$ is related to the $\Lambda_b$  polarization,
 $\alpha_\Lambda$ is the up-down asymmetry parameter of $\Lambda \to p \pi^-$, which can be determined by the experiments~\cite{BESIII:2018cnd}, 
and the definitions of the angles are given in FIG.~\MakeUppercase{\romannumeral 1}\,, in which $\hat{n}_{\Lambda_b}$ is the polarized direction of $\Lambda_b$.
For $b \to s (d)\gamma$,  the longitudinal polarization is defined by
\begin{equation}
P_L \equiv \frac{|H_+^2| - |H_-^2| }{|H_+^2| + |H_-^2|}= -1 +{\cal O}(m_{s(d)}^2/m_b^2)\,,
\end{equation}
where the second equality comes from Eq.~\eqref{equality}. It is interesting to point out that the distribution is independent of $\phi$ due to the angular momentum conservation.
Note that in contrast to ${\bf B}_b \to {\bf B}_n P$ with $P$ a psuedoscalar meson, the up-down asymmetry parameter $\alpha_b$ defined through the equality of
\begin{equation}
\frac{1}{\Gamma}\frac{\partial \Gamma}{\partial \cos \theta}  \propto 1 + P_b\alpha_b \cos\theta = 1 -P_b P_L \cos \theta\,,
\end{equation}
has an opposite sign respecting to $P_L$. It is attributed to that the photon is spin-1 and transversely polarized.

\begin{figure}
		\centering
		\includegraphics[width=0.4\linewidth]{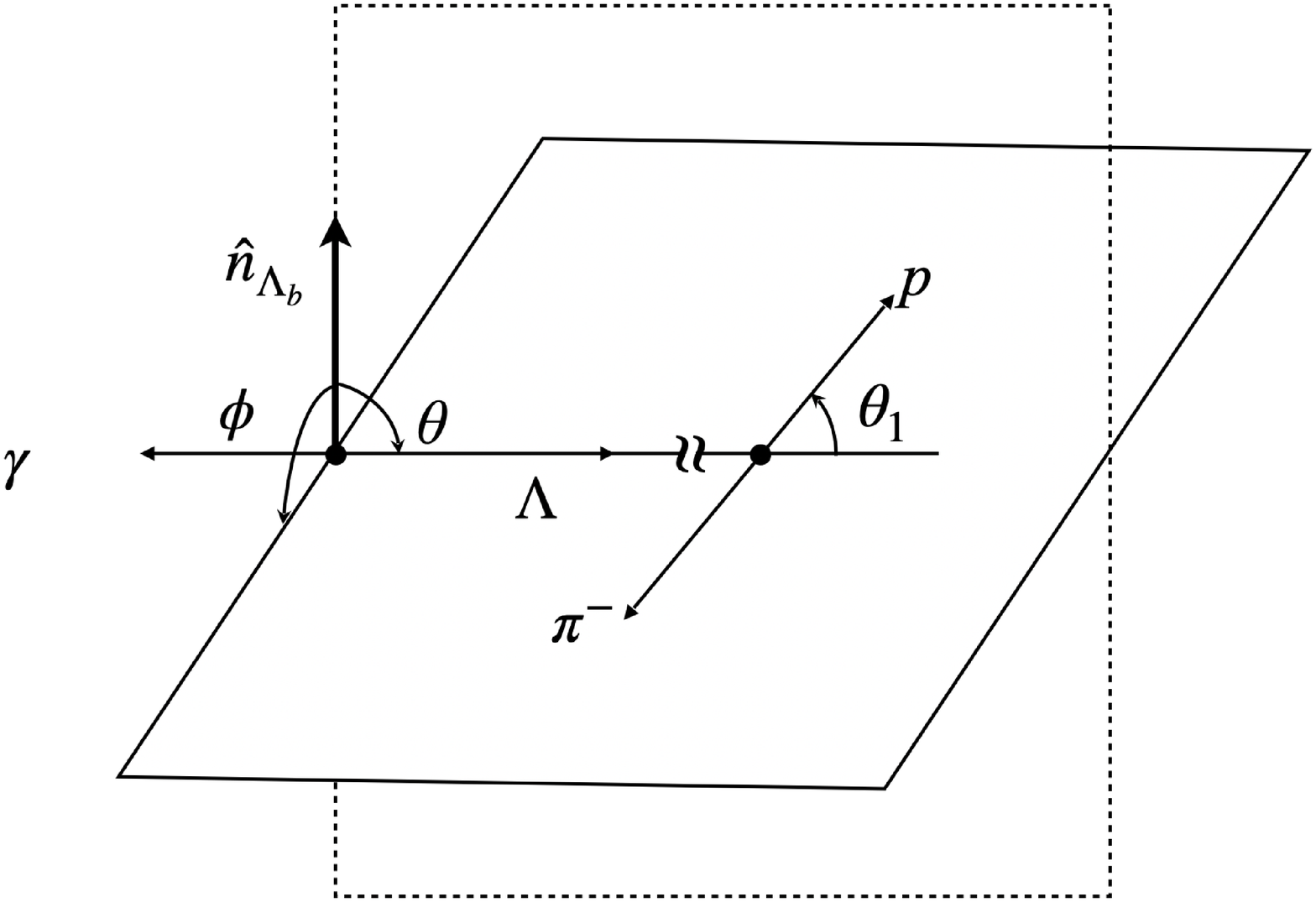}  
		\caption{Angles in ${\cal D}(\vec{ \Omega})$ for $\Lambda_b\to \Lambda(\to p \pi^-)\gamma$.}
\end{figure}

\section{Numerical Results and Discussions}\label{num}

For the CKM  matrix elements, we take the Wolfenstein parametrization,   given as 
\begin{equation}
\lambda=0.22650 \pm 0.00048, \quad A=0.790_{-0.012}^{+0.017}, \quad \rho=0.141_{-0.017}^{+0.016}, \quad \eta=0.357 \pm 0.011,
\end{equation}
where the values and  uncertainties are quoted from the Particle Data Group~\cite{Zyla:2020zbs}.
The values of $\beta_{q_Iq_I'}$ and  $\beta_{Q_b} $ can be found in  Ref.~\cite{Geng:2021nkl}, in which $\beta_{Q_b}$ are taken to be the same for all ${\bf B}_b$ due to the heavy quark symmetry. 
The values of $\beta_{sq_I}$  and $\beta_{ss}$ are taken to be slightly larger than $\beta_{q_Iq_I'}$, since strange quarks are heavier than $u$ and $d$, resulting in a smaller diquark system. 
We note that $\beta_{sq_I}$  and $\beta_{ss}$ are also used for $ \beta_{Q} $ of ${\bf B}_n$. Taking $\Xi^0$ as an example, 
the diquark system is made of $u$ and $s$, so we have $ \beta_{qq'}=\beta_{sq_I} $ and $ \beta_{Q}=\beta_{ss} $.
Our numerical results of the form factors are listed in Table~\ref{tab:ff},
where $ \Xi_b$ and $\Xi $ in the second line stand for either $ \Xi_b^{0} $ and $ \Xi^{0} $, or $ \Xi_b^{-} $ and $ \Xi^{-} $, respectively.

\begin{table}[htbp]
	\centering
	\caption{Theoretical Inputs for the baryon wave functions of the LFQM in the unit of GeV}\label{table1}
	\begin{tabular}{cccccccc}
		\hline
		\hline
		$ m_u $ & $ m_d $ & $ m_s $ & $ m_b $ & $ \beta_{q_Iq_I'} $ & $ \beta_{sq_I} $ & $ \beta_{ss} $ & $ \beta_{Q_b} $ \\
		\hline
		0.26 & 0.26 & 0.31 & 4.88 & 0.365 & 0.373 & 0.377 & 0.601 \\
		\hline
		\hline
	\end{tabular}
\end{table}

\begin{table}[htbp]
	\centering
	\caption{Form Factors of $ {\bf B}_b\to {\bf B}_n\gamma $}
	\label{tab:ff}
	\begin{tabular}{ccccc}
		\hline
		\hline
		Channel &$ f_2^{TV} $ &\vline& Channel &$ f_2^{TV} $\\
		\hline
		$ \Lambda_b \,\to \Lambda \;\gamma  $ & $ -0.123 $ &\vline& $ \Xi_b^0 \;\to \Sigma^0\gamma $ & $ -0.096 $\\
		$ \Xi_b \,\to \Xi \gamma $ & $ 0.143 $ &\vline& $ \Xi_b^- \,\to \Sigma^-\gamma $ & $ -0.134 $\\
		$ \Lambda_b \:\to n \;\:\gamma $ & $ 0.135 $  &\vline& $ \Xi_b^0 \;\,\to \Lambda\;\gamma $ & $ -0.056 $\\
		\hline
		\hline
	\end{tabular}
\end{table}

To check our results in the LFQM, we would also like to study the decays based on  the $SU(3)_F$ flavor symmetry.
With $SU(3)_F$, the wave functions among
the low-lying octet baryons share the identical spacial distribution, so as the wave functions of  the anti-triplet bottom baryons. 
Consequently,  by substituting $s(d)$ for $b $ in $ {\bf B}_b $ and taking the inner products of the spin-flavor wave functions with ${\bf B}_n$, 
the relative sizes of the form factors among $ {\bf B}_b\to {\bf B}_n\gamma $ can be determined. The  $SU(3)_F$ relations~\cite{Wang:2020wxn} are given in Table~\MakeUppercase{\romannumeral 3} with $\lambda_{s(d)} = V_{tb}V^*_{ts(d)}$.
By taking the experimental data for the branching ratio of $ \Lambda_b \to \Lambda \gamma $ as a theoretical input,  we can obtain the $SU(3)_F$ predictions for 
$ {\bf B}_b\to {\bf B}_n\gamma $.

\begin{table}[htbp]
	\centering
	\caption{Amplitude ratios of $ {\bf B}_b\to {\bf B}_n\gamma $ given by $ SU(3)_F $}
	\label{tab:IRA}
	\begin{tabular}{ccccccc}
		\hline
		\hline
		Channel & $ \Lambda_b\to\Lambda\gamma $ & $ \Xi_b\to\Xi\gamma $ & $ \Lambda_b\to n\gamma $ & $ \Xi_b^0\to\Sigma^0\gamma $ & $ \Xi_b^-\to\Sigma^-\gamma $ & $ \Xi_b^0\to\Lambda\gamma $ \\
		\hline
		Amplitude & $ \lambda_s $ & $ -\sqrt{\frac{3}{2}}\lambda_s $ & $ -\sqrt{\frac{3}{2}} \lambda_d $ & $\frac{\sqrt{3}}{2}\lambda_d$ & $ \sqrt{\frac{3}{2}} \lambda_d$ & $ \frac{1}{2}\lambda_d $ \\
		\hline
		\hline
	\end{tabular}
\end{table}

\begin{table}[htbp]
	\centering
	\caption{Numerical results of the branching ratios for $ {\bf B}_b\to {\bf B}_n\gamma $}
	\label{numtab}
	\begin{tabular}{clccccc}
		\hline
		\hline
		Quark level & ~Branching Ratios & ~LFQM~ & $SU(3)_F$ & $SU(3)_F$~\cite{Wang:2020wxn} & 
		Other Models & Exp. Data\\
		\hline
		\multirow{5}*{$ b\to s\gamma $ } 
		& \multirow{3}{*}{$10^{6} \mathcal{B}(\Lambda_b \,\to \Lambda \;\gamma ) $} &  \multirow{3}{*}{~~$ 7.1\pm 0.3 $~~} & \multirow{3}{*}{$ 7.1\pm1.7 $} & \multirow{3}{*}{$ 7.1\pm3.4 $} & ~~$  7.3\pm1.5 $~\cite{Wang:2008sm}~~ & \multirow{3}{*}{$ 7.1\pm1.7 $~\cite{LHCb:2019wwi}} \\[0.1pt]
		& & & & & $ 4.0 $~\cite{Gutsche:2013pp} & \\[0.1pt]
		& & & & & $ 10.0 $~\cite{Faustov:2017wbh} & \\[0.1pt]
		& $10^{5} \mathcal{B}(\Xi_b^0 \,\to \Xi^0 \gamma)  $ & $ 1.0\pm 0.1 $ & $ 1.1\pm0.3 $ &$ 1.16\pm0.60 $& $ 1.02^{+0.60}_{-0.46} $~\cite{Olamaei:2021eyo} & \\
		& $10^{5} \mathcal{B}(\Xi_b^- \to \Xi^- \!\gamma) $ & $ 1.1\pm 0.1 $ & $ 1.2\pm0.3 $ &$ 1.23\pm0.64 $& $ 1.08^{+0.63}_{-0.49} $~\cite{Olamaei:2021eyo}  & $ <13 $~\cite{LHCb:2021hfz} \\
		\hline
		\multirow{5}*{$ b\to d\gamma $ }
		& \multirow{2}{*}{$10^{7} \mathcal{B}(\Lambda_b \:\to n \;\:\gamma) $} & \multirow{2}{*}{$ 4.0\pm 0.4 $} & \multirow{2}{*}{$ 4.9\pm1.2 $} &\multirow{2}{*}{$ 5.03\pm2.67 $} & $ 3.69^{+3.76}_{-1.95} $~\cite{Liu:2019rpm} & \\[0.1pt]
		& & & & & $ 3.7 $~\cite{Faustov:2017ous} & \\[0.1pt]
		& $ 10^{7}\mathcal{B}(\Xi_b^0 \;\to \Sigma^0\gamma) $ & $ 2.1\pm 0.2 $ & $ 2.6\pm0.6 $ &$ 2.71\pm1.50 $&  $ 5.77^{+3.16}_{-2.47} $~\cite{Olamaei:2021eyo} & \\
		& $  10^{7}\mathcal{B}(\Xi_b^- \,\to \Sigma^-\gamma) $ & $ 4.4\pm 0.4 $ & $ 5.5\pm1.3 $ &$ 5.74\pm3.21 $&  $ 6.14^{+3.36}_{-2.63} $~\cite{Olamaei:2021eyo} & \\
		& $ 10^{8}\mathcal{B}(\Xi_b^0 \;\,\to \Lambda\;\gamma) $ & $ 7.4 \pm 0.7 $ & $ 8.7\pm2.1 $ & $ 9.17\pm5.10 $& $ $ & \\
		\hline
		\hline
	\end{tabular}
\end{table}

Our  results of the branching ratios from the LFQM and $SU(3)_F$ 
are given in Table~\ref{numtab}, 
where we have also shown the $SU(3)_F$ evaluations in Ref.~\cite{Wang:2020wxn} and some of other theoretical predictions
in the literature, such as
LCSR~\cite{Wang:2008sm,Olamaei:2021eyo}, BSE~\cite{Liu:2019rpm} and QM~\cite{Gutsche:2013pp,Faustov:2017ous},
as well as the current experimental data~\cite{LHCb:2019wwi,LHCb:2021hfz}.
In particular, for $b \to s \gamma$ in the LFQM approach, we find that
${\cal B}(\Lambda_b\to \Lambda \gamma) = (7.1\pm 0.3)\times 10^{-6}$,
 which agrees well with the experimental measured value.
In addition, we obtain that 
\begin{equation}
{\cal B}(\Xi_b^0 \to \Xi^0 \gamma) = (1.0\pm 0.1)\times 10^{-5}\,,\quad {\cal B}(\Xi_b^- \to \Xi^- \gamma) = (1.1\pm 0.1)\times 10^{-5}\,,
\end{equation}
which are 1.5 times larger than  $ {\cal B}(\Lambda_b\to \Lambda\gamma) $.
Note that  the form factors for $ \Xi_b^0\to\Xi^0\gamma $ and $ \Xi_b^-\to\Xi^-\gamma $ are exactly the same due to the isospin symmetry, but their branching ratios slightly differ due to the lifetime difference.
Similarly,  we have  that
\begin{equation}\label{xie}
\Gamma( \Xi_b^-\to\Sigma^-\gamma)  =  2 \Gamma( \Xi_b^0\to\Sigma^0\gamma )
\end{equation}
guaranteed by the isospin symmetry.

In addition, our  LFQM results  in $ b\to s\gamma $ agree well with both the predictions given by $ SU(3)_F $ and Refs.~\cite{Wang:2008sm,Olamaei:2021eyo,Liu:2019rpm}. Note that the values based on $SU(3)_F$  in Ref.~\cite{Wang:2020wxn} was made within 2$ \sigma $ errors with respect to the experimental result of $ \mathcal{B}(\Lambda_b \to \Lambda \gamma ) $. Furthermore, 
our $SU(3)_F$ results of the center values in Table~\ref{numtab} also slightly differ from those in  Ref.~\cite{Wang:2020wxn}.
These differences arise from  the long-distance contributions of $ {\bf B}_{b}\to {\bf B}_{n}\psi_i(\to\gamma) $ included  in Ref. [22], which modify the ratios between $b\to s \gamma$ and $b\to d \gamma$.

It is interesting to see that the decay branching ratios associated with $ b\to d\gamma $ in  the LFQM are about $20\%$ smaller than those predicted by the $SU(3)_F$ symmetry,
which clearly show the $ SU(3)_F $ breaking effects.
In contrast, the ones given by Ref.~\cite{Olamaei:2021eyo} are larger than the $SU(3)_F$ predicted values.
Note that the equalities of Eqs.~\eqref{feq} and  \eqref{xie} do not hold in Ref.~\cite{Olamaei:2021eyo}\,.
Future experimental searches on $ \Xi_b\to\Sigma\gamma $ could discriminate the various theoretical approaches.

\section{Conclusions}\label{con}

We have performed a systematic analysis of $ {\bf B}_b\to {\bf B}_n\gamma $ based on the LFQM. 
We have obtained ${\cal B}(\Lambda_b\to \Lambda \gamma) = (7.1\pm 0.3)\times 10^{-6}$, ${\cal B}(\Xi_b^0 \to \Xi^0 \gamma) = (1.0\pm 0.1)\times 10^{-5}\,,$ and ${\cal B}(\Xi_b^- \to \Xi^- \gamma) = (1.1\pm 0.1)\times 10^{-5}$. Our results agree with   the current experimental data and are consistent with other theoretical values in the literature.
In addition, for $b\to s \gamma$ we have  found that our results in the LFQM are in good agreement with those based on
the $SU(3)_F$ symmetry. Moreover, we have demonstrated  that the $SU(3)_F$ breaking effects for $b\to d \gamma$ are as large as $20\%$. 

We have also explicitly shown that  $ f_2^{TV} = f_2^{TA} $ at $ k^2=0 $ in the LFQM, resulting in that $P_L= -\alpha = -1 +{\cal O}(m_{s(d)}^2/m_b^2)$ for $b\to s(d) \gamma$, which are independent of the theoretical input. A dedicated experimental measurement of   the angular distribution of $\Lambda_b \to \Lambda (\to p \pi ^ -) \gamma$ 
 are strongly recommended for testing the SM and probing possible effects from new physics.

\begin{acknowledgments}
	We want to thank Dr. Tien-Hsueh Tsai for his valuable assistance on the numerical part of this work. 
\end{acknowledgments}

\newpage
\appendix

\section{Dirac Spinor in Light-Front formalism}\label{appB}

We adopt the notation given  in Ref.~\cite{Zhang:1994ti} for  the light front formalism. The Dirac spinors turn out to be
\begin{equation}
	\begin{aligned}
		&u(p, \lambda)=\frac{1}{\sqrt{p^{+}}}\left(p^{+}+\beta +\boldsymbol{\alpha}_{\perp} \boldsymbol{p}_{\perp}\right) \times \begin{cases}\chi(\uparrow) & \text { for } \lambda=+1 \\
			\chi(\downarrow) & \text { for } \lambda=-1\end{cases}, \\
		&v(p, \lambda)=\frac{1}{\sqrt{p^{+}}}\left(p^{+}-\beta m+\boldsymbol{\alpha}_{\perp} \boldsymbol{p}_{\perp}\right) \times \begin{cases}\chi(\downarrow) & \text { for } \lambda=+1 \\
			\chi(\uparrow) & \text { for } \lambda=-1\end{cases},
	\end{aligned}
\end{equation}
where $ \beta=\gamma^0$ and $\boldsymbol{\alpha}_{\perp}=(\gamma^0\gamma^1,\gamma^0\gamma^2) $
with $\gamma^\alpha$ ($\alpha = 0,1,2,3$) being the Dirac gamma matrices,
while the two $ \chi $-spinors are given by
\begin{equation}
	\chi(\uparrow)=\frac{1}{\sqrt{2}}\left(\begin{array}{l}
		1 \\0 \\1 \\0
	\end{array}\right) \quad \text { and } \quad \chi(\downarrow)=\frac{1}{\sqrt{2}}\left(\begin{array}{c}
		0 \\1 \\0 \\-1
	\end{array}\right),
\end{equation}
resulting in the helicity eigenstates
\begin{equation}
	u(p,+)=\frac{1}{\sqrt{2p^+}}\begin{pmatrix}
		p^++m\\p^1+ip^2\\p^+-m\\p^1+ip^2
	\end{pmatrix}\qq{and}u(p,-)=\frac{1}{\sqrt{2p^+}}\begin{pmatrix}
		-p^1+ip^2\\p^++m\\p^1-ip^2\\-p^++m
	\end{pmatrix}\,.
\end{equation}
Accordingly, the relations in Eq.~\eqref{extract} and~\eqref{O3T} can be verified directly.

\section{Momentum-Spin-Flavor Wave Function}\label{appC}

The momentum-spin-flavor wave functions  for $ {\bf B}_b $ and $ {\bf B}_n $ are given as 
\begin{equation}
	\begin{aligned}
		\ket{\Xi_b^-}&=\frac{1}{\sqrt{6}}\left[\phi_{3} \chi^{\rho 3}(\ket{dsb}-\ket{sdb})+\phi_{2} \chi^{\rho 2}(\ket{dbs}-\ket{sbd})+\phi_{1} \chi^{\rho 1}(\ket{bds}-\ket{bsd})\right],\\
		\ket{\Xi_b^0}&=\frac{1}{\sqrt{6}}\left[\phi_{3} \chi^{\rho 3}(\ket{usb}-\ket{sub})+\phi_{2} \chi^{\rho 2}(\ket{ubs}-\ket{sbu})+\phi_{1} \chi^{\rho 1}(\ket{bus}-\ket{bsu})\right],\\
		\ket{\Lambda_b^0}&=\frac{1}{\sqrt{6}}\left[\phi_{3} \chi^{\rho 3}(\ket{udb}-\ket{dub})+\phi_{2} \chi^{\rho 2}(\ket{ubd}-\ket{dbu})+\phi_{1} \chi^{\rho 1}(\ket{bud}-\ket{bdu})\right],\\
	\end{aligned}
\end{equation}

\begin{equation}
	\begin{aligned}
		\ket{p}&=\frac{1}{\sqrt{3}} \phi\left[\chi^{\lambda3}\ket{uud}+\chi^{\lambda2}\ket{udu}+\chi^{\lambda1}\ket{duu}\right],\\
		\ket{n}&=\frac{1}{\sqrt{3}} \phi\left[\chi^{\lambda3}\ket{ddu}+\chi^{\lambda2}\ket{dud}+\chi^{\lambda1}\ket{udd}\right],\\
		\ket{\Xi^0}&=\frac{1}{\sqrt{3}}\left[\phi_3\chi^{\lambda3}\ket{ssu}+\phi_2\chi^{\lambda2}\ket{sus}+\phi_1\chi^{\lambda1}\ket{uss}\right],\\
		\ket{\Xi^-}&=\frac{1}{\sqrt{3}}\left[\phi_3\chi^{\lambda3}\ket{ssd}+\phi_2\chi^{\lambda2}\ket{sds}+\phi_1\chi^{\lambda1}\ket{dss}\right],\\
		\ket{\Sigma^+}&=\frac{1}{\sqrt{3}}\left[\phi_3\chi^{\lambda3}\ket{uus}+\phi_2\chi^{\lambda2}\ket{usu}+\phi_1\chi^{\lambda1}\ket{suu}\right],\\
		\ket{\Sigma^-}&=\frac{1}{\sqrt{3}}\left[\phi_3\chi^{\lambda3}\ket{dds}+\phi_2\chi^{\lambda2}\ket{dsd}+\phi_1\chi^{\lambda1}\ket{sdd}\right],\\
	\end{aligned}
\end{equation}

\begin{equation}
	\begin{aligned}
		\ket{\Sigma^0}&=\frac{1}{\sqrt{6}}\left[\phi_{3}\chi^{\lambda 3}(\ket{uds}+\ket{dus})+\phi_{2} \chi^{\lambda 2}(\ket{usd}+\ket{dsu})
		+\phi_{1} \chi^{\lambda 1}(\ket{sud}+\ket{sdu})\right],\\
		\ket{\Lambda}&=\frac{1}{\sqrt{6}}\left[\phi_{3} \chi^{\rho 3}(\ket{uds}-\ket{dus})+\phi_{2} \chi^{\rho 2}(\ket{usd}-\ket{dsu})+\phi_{1} \chi^{\rho 1}(\ket{sud}-\ket{sdu})\right].\\
	\end{aligned}
\end{equation}

Here, the spin wave functions are defined as
\begin{equation}
	\begin{aligned}
		&\chi^{\rho3}_\uparrow=\frac{1}{\sqrt{2}}(\ket{\uparrow\downarrow\uparrow}-\ket{\downarrow\uparrow\uparrow}),\quad	\chi^{\lambda3}_\uparrow=\frac{1}{\sqrt{6}}(2\ket{\uparrow\uparrow\downarrow}-\ket{\uparrow\downarrow\uparrow}-\ket{\downarrow\uparrow\uparrow}),\\
		&\chi^{\rho2}_\uparrow=\frac{1}{\sqrt{2}}(\ket{\uparrow\uparrow\downarrow}-\ket{\downarrow\uparrow\uparrow}),\quad	\chi^{\lambda2}_\uparrow=\frac{1}{\sqrt{6}}(2\ket{\uparrow\downarrow\uparrow}-\ket{\uparrow\uparrow\downarrow}-\ket{\downarrow\uparrow\uparrow}),\\
		&\chi^{\rho1}_\uparrow=\frac{1}{\sqrt{2}}(\ket{\uparrow\uparrow\downarrow}-\ket{\uparrow\downarrow\uparrow}),\quad	\chi^{\lambda1}_\uparrow=\frac{1}{\sqrt{6}}(2\ket{\downarrow\uparrow\uparrow}-\ket{\uparrow\downarrow\uparrow}-\ket{\uparrow\uparrow\downarrow}).\\
	\end{aligned}
\end{equation}

The definitions of momentum wave functions $ \phi_{1,2} $ are given as
\begin{equation}
	\phi_{1,2}\equiv \Phi(\xi_{1,2}, q_{1,2\perp}, \eta_{1,2}, Q_{1,2\perp})=\mathcal{N}\sqrt{\frac{\partial q_{1,2z}}{\partial\xi_{1,2}}\frac{\partial Q_{1,2z}}{\partial\eta_{1,2}}}\exp(-\frac{\vec{Q}_{1,2}^2}{2\beta_Q^2}-\frac{\vec{q}_{1,2}\!^{2}}{2\beta_{qq'}^2}),
\end{equation}
\begin{equation}
	\begin{aligned}
		q_{1,2z}&=\frac{\xi_{1,2} M_{1,2}}{2}-\frac{m_{2,3}^2+q_{1,2\perp}^2}{2\xi_{1,2} M_{1,2}},\quad\,\vec{q}_{1,2}\!^{2}=q_{1,2\perp}^2+q_{1,2z}^2,\\
		Q_{1,2z}&=\frac{\eta_{1,2} M}{2}-\frac{m_{1,2}^2+Q_{1,2\perp}^2}{2\eta_{1,2} M},\quad\vec{Q}_{1,2}^2=Q_{1,2\perp}^2+Q_{1,2z}^2,\\
	\end{aligned}
\end{equation}

\end{document}